\documentclass[aps,prc,twocolumn,preprintnumbers,showpacs,floatfix,nofootinbib]%
  {revtex4-1}
\usepackage{amsmath,amsfonts,graphicx}

\newcommand{\dd}{{\rm d}}

\begin{document}

\preprint{BI-TP 2010/045}

\title{Anisotropic flow far from equilibrium}

\author{Nicolas Borghini, Cl\'ement Gombeaud}
\affiliation{Fakult\"at f\"ur Physik, Universit\"at Bielefeld, Postfach 100131,
  D-33501 Bielefeld, Germany}

\date{\today}

\begin{abstract}
We compute analytically the anisotropic flow in an expanding mixture of several
species of relativistic massive particles. 
We find that a single collision per particle in average already leads to sizable
elliptic flow, with mass ordering between the species. 
\end{abstract}


\maketitle

\section{Introduction}

The hundreds or even thousands of particles in the final state of a 
nucleus--nucleus collision at high energy are emitted asymmetrically in the 
plane transverse to the collision axis. 
This phenomenon, referred to as anisotropic flow, is conveniently quantified by
introducing the Fourier expansion of the (invariant) particle yield~\cite{%
  Voloshin:1994mz}. 
Restricting ourselves to the transverse momentum distribution, 
\begin{subequations}
\label{d2N/d2pT_vn}
\begin{equation}
\label{d2N/dpT}
\frac{\dd^2N}{\dd^2{\bf p}_T} = \frac{1}{2\pi}\frac{\dd N}{p_T\,\dd p_T}
  \bigg[ 1 + \sum_{n=1}^\infty 2v_n(p_T)\cos n\varphi\bigg],
\end{equation}
where the Fourier coefficients are given by 
\begin{equation}
\label{vn}
v_n(p_T) = \frac{\displaystyle\int\!\dd\varphi\,\frac{\dd^2N}{\dd^2{\bf p}_T}\,
  \cos n\varphi}{\displaystyle\int\!\dd\varphi\,\frac{\dd^2N}{\dd^2{\bf p}_T}}.
\end{equation}
\end{subequations}
In Eqs.~\eqref{d2N/d2pT_vn} $\varphi$ denotes the azimuth of outgoing particles 
with respect to the direction of the impact parameter.
Anisotropic flow is thus a {\em collective\/} effect, that signals a correlation
of each particle to the reaction plane. 

Given this collectivity property, a natural modeling of anisotropic flow 
consists in considering the particles as a continuous medium that evolves
according to the laws of relativistic fluid dynamics~\cite{hydrorefs}.
This description implicitly assumes that particles undergo many collisions, to 
ensure (quasi-)local equilibrium. 

On the opposite, one may also investigate anisotropic flow far from equilibrium,
when outgoing particles interact seldom, which is our purpose in this article.
A small number of interactions per particle is expected in several cases. 
The most obvious one is that of peripheral collisions, where only relatively few
particles are emitted: then the number of their reinteractions, integrated over
the whole expansion, will be small. 
Other instances of presumably far-from-equilibrium evolution are the early and 
late stages of the expansion in collisions with arbitrary centrality: 
either {\em before\/} the systems equilibrates, as was considered in 
Ref.~\cite{Krasnitz:2002ng,Broniowski:2008qk} (here, the rate of collisions 
remains small for a short while, before becoming large); 
or at the very end of the evolution, when collisions become seldom around the
kinetic freeze-out. 

In this article, we describe the expanding system as a mixture of several 
relativistic massive components governed by the Boltzmann equation, which we 
solve analytically in the limit of a small number of collisions per particle, 
focusing on calculating the flow harmonics~\eqref{vn}.
To our knowledge, the only previous attempt in that direction is that of 
Ref.~\cite{Heiselberg:1998es}, which however only deals with non-relativistic 
particles in the low-eccentricity limit. 
Here our results will hold across all centralities, and we are not restricted 
to low momenta. 

In Section~\ref{s:model} we introduce our model and outline the computation of 
the average number of collisions per particle and of the resulting anisotropic 
flow. 
Section~\ref{s:computations} contains the technical details of our calculations
in the case of massive particles --- the limiting cases involving massless 
particles are dealt with in Appendices~\ref{s:massless} and \ref{s:Lorentz}. 
We present our results in Section~\ref{s:results} and discuss their meaning, as 
well as the limitations of our model in Section~\ref{s:discussion}. 
Eventually, Appendix~\ref{s:Lauricella} introduces for the sake of reference 
some of the properties of the special function in terms of which we express our 
analytical results.

\section{Model}
\label{s:model}

Our purpose in the following is to describe the evolution of the expanding 
system created in a high-energy nuclear collision in the limit of a small 
average number of collisions per outgoing particle.
For that, we model the system as a two-dimensional mixture of various components
with respective particle masses $m_i$, which can interact elastically with each 
other. 
As we are interested in the anisotropies in the {\em transverse\/} expansion, we
restrict ourselves to the transverse $(x,y)$-plane, neglecting the influence of 
the longitudinal expansion. 
This is a valid ansatz as long as we consider the qualitative aspects of our 
final results only. 
Our system will be ``homogeneous'', i.e. when we consider a thermal-like 
momentum distribution with inverse slope parameter $T$, then the latter will be 
the same everywhere. 
Eventually, we assume that the system is invariant under the parity 
transformation ${\bf x}\to-{\bf x}$ in position space, and that this symmetry 
holds during the whole evolution.

A suitable approach consists in using a Boltzmann description, approximating its
exact solution as a small modification to the solution of the collisionless 
equation. 
To each component of the mixture we associate its distribution function 
$f_i(t,{\bf x},{\bf p}_i)$, normalized to the particle number $N_i$.
The corresponding particle density is
\begin{equation}
\label{n_i}
n_i(t,{\bf x})=\!\int\!\dd^2{\bf p}_i\, f_i(t,{\bf x},{\bf p}_i),
\end{equation}
and the (transverse) momentum distribution
\begin{equation}
\label{dN_i/dpT}
\frac{\dd^2N_i}{\dd^2{\bf p}_i}(t,{\bf p}_i)=\!\int\!\dd^2{\bf x}\, 
  f_i(t,{\bf x},{\bf p}_i),
\end{equation}
Multiplying this momentum distribution by $\cos n\varphi_i$, with $\varphi_i$ 
the azimuth of ${\bf p}_i$, and averaging over $\varphi_i$, one obtains the 
flow coefficient $v_n(p_i,t)$ at time $t$ [Eq.~\eqref{vn}].
The ``usual'' $v_n(p_i)$ is the large-time limit of $v_n(p_i,t)$.

The distribution function $f_i\equiv f_i(t,{\bf x},{\bf p}_i)$ for each 
component obeys the classical relativistic Boltzmann equation\footnote{We 
  neglect any long-range interaction between particles, which could be 
  implemented as a mean-field term, as considered in the context of anisotropic
  flow in Ref.~\cite{Koch:2009wk}.}
\begin{align}
\label{Boltzmann}
\frac{\partial f_i}{\partial t} + {\bf v}_i\cdot\nabla_{{\bf x}}f_i = & \cr
 \sum_k \bigg(1-\frac{1}{2}\delta_{ik}\bigg)&\!\int\!\dd^2{\bf p}_k\,\dd\Theta\,
   (f'_i f'_k-f_i f_k)\,v_{ik}\frac{\dd\sigma_{ik}}{\dd\Theta},\quad
\end{align}
with $f_k\equiv f_k(t,{\bf x},{\bf p}_k)$, 
$f'_i\equiv f_i(t,{\bf x},{\bf p}'_i)$, 
$f'_k\equiv f_k(t,{\bf x},{\bf p}'_k)$, while the factor 
$1-\frac{1}{2}\delta_{ik}$ ensures that the equation holds for both identical 
and nonidentical particles. 
The relative velocity is given by
\begin{equation}
\label{v_Moller}
v_{ik}=\sqrt{({\bf v}_i-{\bf v}_k)^2-\frac{({\bf v}_i\times{\bf v}_k)^2}{c^2}},
\end{equation}
and $\dd\sigma_{ik}/\dd\Theta$ is a differential scattering cross section for 
collisions between particles of species $i$ and $k$.
Hereafter we shall consider only elastic collisions, so that the number of 
particles for each species remains constant, with an isotropic, constant 
differential cross section, which we shall for the sake of brevity denote 
$\sigma_{\rm d}$.
Note that in our two-dimensional approach the cross section has the dimension of
a length. 

Integrating the Boltzmann equation over ${\bf x}$ yields for the first term in 
the left-hand side the time derivative of the momentum distribution~\eqref{%
  dN_i/dpT}. 
On the other hand, given the assumed parity of the system, the spatial gradient 
$\nabla_{{\bf x}}f_i$ is odd under the parity transformation, and its integral 
over ${\bf x}$ vanishes. 
The integrated Boltzmann equation thus describes the time evolution of the 
momentum distribution, in particular of its anisotropies, under the influence 
of elastic collisions

In the absence of rescatterings, the Boltzmann equation is satisfied by the 
free-streaming solutions
\begin{equation}
\label{free-stream_sol}
f_i^{(0)}(t,{\bf x},{\bf p}_i) = f_i^{(0)}(0,{\bf x}-{\bf v}_i t,{\bf p}_i), 
\end{equation}
which only depend on the initial distribution at $t=0$. 
If the latter is isotropic in momentum space, the free-streaming solution 
remains isotropic, i.e.\ it does not develop any anisotropic flow. 
As initial condition, we shall consider a factorized distribution, with a 
Gaussian dependence in position space and an isotropic distribution 
$\tilde{f}_i(p_i)$ in momentum space:
\begin{equation}
\label{dist(t=0)}
f_i^{(0)}(0,{\bf x},{\bf p}_i) = \frac{N_i}{4\pi^2 R_x R_y}\tilde{f}_i(p_i)
  \exp\!\bigg(\!-\frac{x^2}{2R_x^2}-\frac{y^2}{2R_y^2}\bigg),
\end{equation}
with $p_i=|{\bf p}_i|$ and the normalization 
$\int\!\dd p_i\,p_i\tilde{f}_i(p_i)=1$, and $R_y>R_x$, i.e. the reaction plane 
lies along the $x$-axis. 
Note that this distribution is invariant under the ${\bf x}\to-{\bf x}$ 
transformation, and that with the assumed non-parity-violating cross section the
property holds throughout the evolution.

Hereafter we shall rewrite the mean square radii as
\begin{subequations}
\begin{equation}
\label{Rx,Ry}
R_x^2=\frac{R^2}{1+\epsilon},\quad R_y^2=\frac{R^2}{1-\epsilon},
\end{equation}
that is
\begin{equation}
\label{eccentricity}
R=\frac{1}{\sqrt{2}}\bigg(\frac{1}{R_x^2}+\frac{1}{R_y^2}\bigg)^{\!\!-1/2},\  
  \epsilon=\frac{R_y^2-R_x^2}{R_x^2+R_y^2}.
\end{equation}
\end{subequations}
$\epsilon$ is thus the usual eccentricity of the initial distribution.

When collisions are present, the distribution function, starting with the same 
initial condition at $t=0$, changes to 
\begin{equation}
\label{dev-f_i}
f_i(t,{\bf x},{\bf p}_i) =
  f_i^{(0)}(t,{\bf x},{\bf p}_i)+f_i^{(1)}(t,{\bf x},{\bf p}_i)+\cdots,
\end{equation}
with $f_i^{(1)}$ much smaller than $f_i^{(0)}$.\footnote{The ``small parameter''
  that controls the expansion is the average number of collisions per particle, 
  which is proportional to $\sigma_{\rm d}$.}
Due to the collisions, $f_i$ may develop anisotropies in momentum space, which, 
neglecting the next terms in the expansion, will be those of $f_i^{(1)}$. 
Reinserting expansion~\eqref{dev-f_i} in the Boltzmann equation, one should 
as a first approximation consider only $f_i^{(0)}+f_i^{(1)}$ in the left-hand 
side and $f_i^{(0)}$ in the right-hand side.
Our approach thus differs from the traditional Chapman-Enskog formulation, in 
which the collision integral of the leading term in the expansion 
vanishes~\cite[Chapter 6]{Huang:statphys}.

Let us now focus on the momentum anisotropies.
The gain term in the collision integral does not contribute to developing such 
anisotropies, at least to the assumed order of approximation, thanks to the 
isotropy of the cross section. 
More precisely, the gain term involves the {\em free-streaming\/} particle 
distributions {\em after\/} the collision.
The ``memory'' of the latter only extends back to the time of their last 
collision --- technically, this amounts to replacing $t=0$ (resp.\ $t$) by the 
collision time $t_{\rm coll}$ (resp.\ by $t-t_{\rm coll}$) in the right-hand 
side of Eq.~\eqref{free-stream_sol} --- and thus bears no correlation to the 
initial geometry --- which means that the product 
$f^{(0)\prime}_i f^{(0)\prime}_k$ does not involve the azimuths $\varphi_i$ and 
$\varphi_k$ of the momenta, so that the integral of 
$f^{(0)\prime}_i f^{(0)\prime}_k v_{ik}\cos n\varphi_i$ over both azimuths 
gives 0.

To derive the numerator of Eq.~\eqref{vn}, we thus just have to consider the 
loss term, proportional to $-f_i^{(0)} f_k^{(0)}v_{ik} \sigma_{\rm d}$.
Let us first not perform the integration over ${\bf p}_k$: the term then depends
only on the velocities ${\bf v}_i$, ${\bf v}_k$ of the colliding particles. 
Multiplying it by $\cos n\varphi_i$, we can then integrate over time and the 
position, as well as over the azimuths $\varphi_k$ and $\varphi_i$.
This yields a function, hereafter referred to as the ``anisotropy of the 
velocity distribution'', which integrated over $|{\bf p}_k|$ yields the 
numerator in the definition~\eqref{vn} of the flow harmonic $v_n(p_i)$ for 
particle $i$.

Anticipating on one of our results, this distribution involves the product of 
$\tilde{f}_i(p_i)\tilde{f}_k(p_k)$ with a function --- which we compute for 
$n=2$ and $n=4$ in Section~\ref{s:computations} --- that depends on 
eccentricity, $|{\bf v}_i|$ and $|{\bf v}_k|$; 
this function can however not depend on the masses $m_i$, $m_k$ of the 
particles, since the latter are decoupled from the velocities due to the 
factorization of the space and momentum parts in the initial 
distribution~\eqref{dist(t=0)}.
This means that the intrinsic dependence of the Fourier coefficient $v_n(p_i)$ 
is on the velocity $|{\bf v}_i|$, rather than on the momentum $p_i$:
when comparing coefficients of different particle species, they take the same 
value at fixed velocity, i.e.\ as a function of momentum they exhibit mass 
ordering.
In order to emphasize the independence of the (qualitative) mass ordering of 
flow from the assumed initial spectrum, we shall present in Section~\ref{%
  s:results} results obtained with two different choices for $\tilde{f}(p)$. 

Once the anisotropies of the velocity distribution have been obtained, we can 
assume a specific form for the initial momentum distribution, and compute the 
flow coefficients $v_n(p_i)$. 
The denominator in definition~\eqref{vn} is to our order of approximation given 
by replacing $f_i$ by $f^{(0)}_i$ in the integral~\eqref{dN_i/dpT}, which yields
\begin{equation}
\label{denom-vn}
\int\!\dd\varphi_i\,\frac{\dd^2N_i}{\dd^2{\bf p}_i} = N_i\tilde{f}_i(p_i).
\end{equation}

To ensure the internal consistency of our model, the average number of 
collisions per outgoing particle in the evolution should be small, say at most 
equal to 1.
This number of collisions is quite naturally the integral over time of the 
collision rate, which for two beams of particle densities $n_i$, $n_k$ with 
velocities ${\bf v}_i$, ${\bf v}_k$ reads
\begin{equation}
\label{collision_rate}
\frac{\dd N_{\rm coll}}{\dd t\,\dd^2\bf x} = \int\!\dd\Theta\,
  \frac{\dd\sigma_{ik}}{\dd\Theta}\,n_i n_k v_{ik},
\end{equation}
with the densities given by Eq.~\eqref{n_i} and the relative velocity by 
Eq.~\eqref{v_Moller}.
One recognize here (minus) the loss term of the Boltzmann equation~\eqref{%
  Boltzmann}, integrated over ${\bf p}_i$.
The calculation is thus very similar to that of the flow coefficients, with the
only difference that there is no $\cos n\varphi_i$ term --- which amounts to 
taking $n=0$ --- and that there is a final integration over $|{\bf p}_i|$ --- 
which is the equivalent of computing the ``integrated flow''. 
Fixing the average number of collisions to 1 will of course be equivalent to 
fixing the value of the cross section $\sigma_{\rm d}$ (or rather, of the 
dimensionless quantity $\sigma_{\rm d}/R$), which allows us to present numerical
values for $v_2(p_i)$ and $v_4(p_i)$ in Section~\ref{s:results}.

\section{Explicit calculations}
\label{s:computations}

In this Section we detail our computations of the anisotropy of the velocity 
distribution and of the mean number of collisions per particle, whose principle 
was presented in the previous Section. 
The reader who is not interested in technical issues but is willing to trust us
may skip directly to Section~\ref{s:results}, which only relies on Eqs.~\eqref{%
  limvp_0} and~\eqref{def-F}.

More specifically, we shall derive the anisotropy of the distribution in 
${\bf v}_i$ for particles of species $i$ induced by collisions with particles 
of a different species $k$. 
We shall arbitrarily call the former ``diffusing particles'' and the latter 
``scattering centers''.
In the most general case where both types are massive, the relative 
velocity~\eqref{v_Moller} can be rewritten after some simple algebra as
\begin{equation}
\label{v_ik(mass)}
v_{ik} = c\sqrt{\big[1\!-\!\beta_i\beta_k\cos(\varphi_i\!-\!\varphi_k)\big]^2 -
  (1\!-\!\beta_i^2)(1\!-\!\beta_k^2)},
\end{equation}
where $\beta_i\equiv |{\bf v}_i|/c$, $\beta_k\equiv |{\bf v}_k|/c$.
Note that this expression is simpler when one of the $\beta$ is equal to 1, 
i.e.\ for massless particles. 
Accordingly the corresponding calculations, which we present in 
Appendices~\ref{s:massless} and \ref{s:Lorentz}, become much more 
straightforward.

As argued below Eq.~\eqref{dev-f_i}, we need only consider the loss term of the
collision integral, computed with the free-streaming solution~\eqref{%
  free-stream_sol} and integrated over ${\bf x}$.
The starting point of the calculation is thus the kernel
\begin{widetext}
\begin{align}
\label{kernel}
{\cal C}(t,{\bf p}_i,{\bf p}_k) \equiv &-\int\!\dd^2{\bf x}\,\dd\Theta\,
   f^{(0)}_i(t,{\bf x},{\bf p}_k) f^{(0)}_k(t,{\bf x},{\bf p}_k)\,v_{ik}
  \sigma_{\rm d} \cr
 = &-\frac{N_iN_k\sigma_{\rm d}c\sqrt{1-\epsilon^2}}{8\pi^2 R^2}\,
  \tilde{f}_i(p_i)\,\tilde{f}_k(p_k)\,
    \sqrt{\big[1-\beta_i\beta_k\cos(\varphi_i\!-\!\varphi_k)\big]^2 -
    (1-\beta_i^2)(1-\beta_k^2)}\cr
   & \quad\times\exp\!\bigg[\!-\frac{c^2t^2}{4R^2}\Big(
      \big[1+\epsilon\cos 2\varphi_i\big]\beta_i^2 - 
      2\big[\epsilon\cos 2(\varphi_i\!+\!\varphi_k) +
         \cos 2(\varphi_i\!-\!\varphi_k)\big]\beta_i\beta_k + 
      \big[ 1+\epsilon\cos 2\varphi_k\big]\beta_k^2\Big) \bigg],\qquad\,
\end{align}
where we have performed the Gaussian integral over ${\bf x}$ and that over the 
``solid'' angle $\Theta$.

As explained in Section~\ref{s:model}, this kernel allows us to compute the 
number of collisions --- by integrating it over time, $\varphi_i$, $\varphi_k$, 
$|{\bf p}_k|$ and $|{\bf p}_i|$ --- as well as the anisotropic flow coefficients
$v_n(p_i)$ --- by multiplying with $\cos n\varphi_i$ before integrating over 
$\varphi_i$, and leaving out the final integration over $|{\bf p}_i|$. 

The first integral, over time ranging from 0 to $\infty$, is trivial
\begin{align}
\label{int_t(kernel)}
\int_0^\infty\!\dd t\,{\cal C}(t,{\bf p}_i,{\bf p}_k) = &
  -\frac{N_iN_k\sigma_{\rm d}\sqrt{1-\epsilon^2}}{8\pi^{3/2}R}\,
  \tilde{f}_i(p_i)\,\tilde{f}_k(p_k) \cr
  & \quad\times
      \Bigg(\frac{\big[1-\beta_i\beta_k\cos(\varphi_i\!-\!\varphi_k)\big]^2 -
        (1-\beta_i^2)(1-\beta_k^2)}{
        \big[1+\epsilon\cos 2\varphi_i\big]\beta_i^2 - 
        2\big[\epsilon\cos 2(\varphi_i\!+\!\varphi_k) +
          \cos 2(\varphi_i\!-\!\varphi_k)\big]\beta_i\beta_k + 
        \big[ 1+\epsilon\cos 2\varphi_k\big]\beta_k^2} \Bigg)^{\!\!1/2},\qquad
\end{align}
which then has to be multiplied by $\cos n\varphi_i$, possibly with $n=0$ for 
the average number of collisions.

A convenient approach to computing the angular integrals is to perform the 
change of variables 
\begin{equation}
\label{phi-vs-alpha,delta}
(\varphi_i,\varphi_k) \to \Big( \phi^{\rm p}=\frac{\varphi_i\!+\!\varphi_k}{2},
  \ \delta=\frac{\varphi_i\!-\!\varphi_k}{2} \Big),
\end{equation}
which involves a Jacobian that can be reabsorbed in the limits of integration.
The angular part to be integrated then reads
\begin{equation}
\label{limvppp}
\frac{\cos n(\phi^{\rm p}+\delta)
  \sqrt{\big(1-\beta_i\beta_k\cos 2\delta\big)^2 - (1-\beta_i^2)(1-\beta_k^2)}}%
  {\sqrt{a_\delta\cos 2\phi^{\rm p} + b_\delta\sin 2\phi^{\rm p} + c_\delta}},
\end{equation}
with $\delta$-dependent coefficients
$a_\delta\equiv[(\beta_i^2+\beta_k^2)\cos 2\delta-2\beta_i\beta_k]\epsilon$,
$b_\delta\equiv(\beta_k^2-\beta_i^2)\epsilon\sin 2\delta$, and
$c_\delta\equiv\beta_i^2+\beta_k^2-2\beta_i\beta_k\cos 2\delta$.

The simplest is to perform first the integration over $\phi^{\rm p}$, which 
gives
\begin{equation}
\label{limvpp}
\int_{-\pi}^{\pi}\!\frac{\cos n(\phi^{\rm p}+\delta)\,\dd\phi^{\rm p}}{%
  \sqrt{a_\delta\cos 2\phi^{\rm p} + b_\delta\sin 2\phi^{\rm p} + c_\delta }} = 
  \begin{cases}
    \dfrac{4}{\sqrt{1+\epsilon}}\,
     K\!\bigg(\!\sqrt{\dfrac{2\epsilon}{1+\epsilon}}\bigg)
     \dfrac{1}{\sqrt{\beta_i^2-2\beta_i\beta_k\cos 2\delta + \beta_k^2}} & 
     \text{ for }n=0, \\
     \dfrac{\pi\epsilon}{2}\,
     {}_2F_1\!\big(\frac{3}{4},\frac{5}{4};2;\epsilon^2\big)\,
     \dfrac{\beta_i^2-2\beta_i\beta_k\cos 2\delta + \beta_k^2\cos 4\delta}%
     {(\beta_i^2-2\beta_i\beta_k\cos 2\delta + \beta_k^2)^{3/2}} & 
     \text{ for }n=2, \\
     -\dfrac{3\pi\epsilon^2}{16}\,
     {}_2F_1\!\big(\frac{5}{4},\frac{7}{4};3;\epsilon^2\big)\\
     \qquad\quad\times 
     \dfrac{\beta_i^4-4\beta_i^3\beta_k\cos 2\delta + 
       6\beta_i^2\beta_k^2\cos 4\delta - 4\beta_i\beta_k^3\cos 6\delta + 
       \beta_k^4\cos 8\delta}%
     {(\beta_i^2-2\beta_i\beta_k\cos 2\delta + \beta_k^2)^{5/2}}\!\!\!\! & 
     \text{ for }n=4,
  \end{cases}
  \end{equation}
\end{widetext}
where $K$ is the complete elliptic integral of the first kind and ${}_2F_1$ is 
the Gaussian hypergeometric function.\footnote{The hypergeometric functions in 
  the second and third line can also be rewritten in terms of combinations of 
  powers of $\epsilon$ and of complete elliptic integrals $K$ and $E$ of 
  $\sqrt{2\epsilon/(1+\epsilon)}$. 
  The formulations given in Eq.~\eqref{limvpp} are more compact, and allow one 
  to see more easily the small-$\epsilon$ behavior.
  Reciprocally, one might rewrite the expression for $n=0$ using the identity
  \[
  \frac{4}{\sqrt{1+\epsilon}}
    K\!\bigg(\!\sqrt{\dfrac{2\epsilon}{1+\epsilon}}\bigg) = 
  2\pi\,{}_2F_1\!\bigg(\frac{1}{4},\frac{3}{4};1;\epsilon^2\bigg).
  \]} 

The next integral, over $\delta\in[-\pi,\pi]$, is tedious, yet feasible 
analytically. 
A first step is to factorize the terms that come with half-integer powers in 
a rational fraction of $\cos^2\delta$:
\begin{align}
\sqrt{\frac{%
  \big(1-\beta_i\beta_k\cos 2\delta\big)^2 - (1-\beta_i^2)(1-\beta_k^2)}%
  {(\beta_i^2-2\beta_i\beta_k\cos 2\delta + \beta_k^2)^{n+1}}} =\qquad &\cr
  \sqrt{\frac{(1-x_1\cos^2\delta)(1-x_2\cos^2\delta)}%
    {(1-x_3\cos^2\delta)^{n+1}}}, &
\end{align}
with 
\begin{align}
x_1 &= \dfrac{2\beta_i\beta_k\Big[1+\beta_i\beta_k-
  \sqrt{(1-\beta_i^2)(1-\beta_k^2)}\Big]}{(\beta_i+\beta_k)^2}, \cr
x_2 &= \dfrac{2\beta_i\beta_k\Big[1+\beta_i\beta_k+
  \sqrt{(1-\beta_i^2)(1-\beta_k^2)}\Big]}{(\beta_i+\beta_k)^2}, \\
x_3 &= \dfrac{4\beta_i\beta_k}{(\beta_i+\beta_k)^2} = 
  \frac{x_1x_2}{\beta_i\beta_k}. \nonumber
\end{align}

Next, one performs the change of variable $u=\cos^2\delta$, which yields extra
terms in the denominator under the square root: 
the resulting integrals are of the form 
\begin{equation}
\label{limvpp2}
\int_0^1\!\dd u\,{\cal P}_n(u)\,\sqrt{\frac{(1-x_1u)(1-x_2u)}%
  {u(1-u)(1-x_3u)^{n+1}}},
\end{equation}
with ${\cal P}_n$ a polynomial of degree $n$, coming from the numerators in 
Eq.~\eqref{limvpp}.
Note that the integral over $\delta\in[-\pi,\pi]$ gives 4 times the integral 
over $u\in[0,1]$, yet there comes a factor $\frac{1}{2}$ from the change of 
variables, which partly compensates for it.

In the case $n=0$, the polynomial is a constant, so that we are left with a 
hyperelliptic integral. 
The latter can be expressed in terms of the Lauricella hypergeometric function
of three variables $F_D^{(3)}$, which is discussed in Appendix~\ref{%
  s:Lauricella}.
Restoring the prefactors from Eqs.~\eqref{int_t(kernel)} and \eqref{limvpp}, 
and dropping the minus sign which is irrelevant here, one finds
\begin{align}
\label{limvp_0}
\frac{N_iN_k\sigma_{\rm d}}{\sqrt{\pi}R}\,\tilde{f}_i(p_i)\,\tilde{f}_k(p_k)\,
  \sqrt{1-\epsilon}\,K\!\bigg(\!\sqrt{\dfrac{2\epsilon}{1+\epsilon}}\bigg)
  \qquad\quad \cr
  F_D^{(3)}\!\bigg(\frac{1}{2}, \frac{-1}{2}, \frac{-1}{2}, \frac{1}{2}, 1;
    x_1,x_2,x_3\bigg).
\end{align}
To obtain the total number of collisions between particles of species $i$ and 
$k$, one has to assume some shape for the initial spectra $\tilde{f}_i(p_i)$, 
$\tilde{f}_k(p_k)$ and then to integrate over $p_i=mc\beta_i/\sqrt{1-\beta_i^2}$
and $p_k$. 
Given expression~\eqref{limvp_0}, in which the velocities enter through the 
variables $x_1$, $x_2$, $x_3$, an analytical calculation seems highly 
improbable.\footnote{Unless the integral over $p_k$ and/or $p_i$ can be computed
  earlier in the calculation, before some of the other integrals. 
  We have not investigated this possibility.}
Nevertheless, one can use the fact that the Lauricella function in the second 
line actually only takes values between $2/\pi$ and 1, when $\beta_i$ and 
$\beta_k$ vary between 0 and 1, together with the normalization of $\tilde{f}_i$
and $\tilde{f}_k$, to set bounds on the number of collisions at a given 
eccentricity:
\begin{align*}
\frac{2N_iN_k\sigma_{\rm d}}{\pi^{3/2}R}\,\sqrt{1-\epsilon}\,
  K\!\bigg(\!\sqrt{\dfrac{2\epsilon}{1+\epsilon}}\bigg) \leq N_{\rm coll} \leq 
  \qquad\qquad\qquad\cr
\frac{N_iN_k\sigma_{\rm d}}{\sqrt{\pi}R}\,
  \sqrt{1-\epsilon}\,K\!\bigg(\!\sqrt{\dfrac{2\epsilon}{1+\epsilon}}\bigg).
\end{align*}
Dividing by $N_i$ then yields the average number of collisions per particle of 
type $i$.
If one wants to ensure that there is at most one collision per particle, then 
the first inequality translates into a bound on $\sigma_{\rm d}$.
As the eccentricity-dependent part decreases with $\epsilon$ from $\pi/2$ to 0, 
one can fix this bound to ensure less than one collision per particle over the 
whole eccentricity range: 
$\sigma_{\rm d}^{\max} = \sqrt{\pi}R/N_k$.

For the cases $n=2$ and $n=4$, a convenient trick is to rewrite ${\cal P}_n(u)$ 
as a polynomial in $(1-x_3u)$, as e.g.
\begin{align*}
{\cal P}_2(u) = \dfrac{1}{\beta_i^2}\bigg[& 
  \dfrac{(\beta_i+\beta_k)^2}{2}(1-x_3u)^2 \\ 
  &- \beta_k^2(1-x_3u) + 
  \dfrac{(\beta_i+\beta_k)^2}{2} \bigg],\qquad
\end{align*}
which allows us to express the result of the integrals as sums of Lauricella 
functions of three variables only. 
Including the prefactors from Eqs.~\eqref{int_t(kernel)} and \eqref{limvpp}, 
one finds
\begin{subequations}
  \label{def-F}
  \begin{align}
  \int\!\dd t\,\dd\varphi_i\,\dd\varphi_k\,{\cal C}(t,{\bf p}_i,{\bf p}_k)
    \cos n\varphi_i = \qquad\qquad\qquad& \cr
  N_i\tilde{f}_i(p_i)\,N_k\tilde{f}_k(p_k)\,
    {\cal N}_n{\cal K}_n(\epsilon){\cal F}_n(\beta_i,\beta_k),
  \end{align}
  with 
  \begin{align}
  {\cal N}_2 = &\ \frac{\sigma_{\rm d}\sqrt{\pi}}{8R}, \qquad
  {\cal N}_4 = -\frac{3\sigma_{\rm d}\sqrt{\pi}}{64R}, \label{Nn} \\
  {\cal K}_2(\epsilon) &= \sqrt{1-\epsilon^2}\,
     {}_2F_1\!\bigg(\frac{3}{4},\frac{5}{4};2;\epsilon^2\bigg)\,\epsilon 
  \label{K2}\\
  {\cal K}_4(\epsilon) &= \sqrt{1-\epsilon^2}\,
     {}_2F_1\!\bigg(\frac{5}{4},\frac{7}{4};3;\epsilon^2\bigg)\,\epsilon^2,
  \label{K4}
  \end{align}
  and
\begin{widetext}
  \begin{align}
  {\cal F}_2(\beta_i,\beta_k) &= \dfrac{1}{\beta_i^2}\bigg[
    \frac{(\beta_i+\beta_k)^2}{2}\,F_D^{(3)}\!\bigg(\frac{1}{2}, 
      \frac{-1}{2}, \frac{-1}{2}, \frac{-1}{2}, 1; x_1,x_2,x_3\bigg)
    - \beta_k^2\,F_D^{(3)}\!\bigg(\frac{1}{2}, \frac{-1}{2}, \frac{-1}{2},
        \frac{1}{2}, 1; x_1,x_2,x_3\bigg)\qquad\qquad\qquad \cr
   & \qquad\quad + \dfrac{(\beta_i-\beta_k)^2}{2}\,
     F_D^{(3)}\!\bigg(\frac{1}{2}, \frac{-1}{2}, \frac{-1}{2}, \frac{3}{2}, 1; 
       x_1,x_2,x_3\bigg) \bigg],\qquad \label{F2}
  \end{align}
  \begin{align}
  {\cal F}_4(\beta_i,\beta_k) &= \dfrac{1}{\beta_i^4}\bigg[
     \frac{(\beta_i+\beta_k)^4}{2}\,F_D^{(3)}\!\bigg(\frac{1}{2}, \frac{-1}{2}, 
       \frac{-1}{2}, \frac{-3}{2}, 1; x_1,x_2,x_3\bigg) 
     - 2\beta_k^2(\beta_i+\beta_k)^2\,F_D^{(3)}\!\bigg(\frac{1}{2}, 
       \frac{-1}{2}, \frac{-1}{2}, \frac{-1}{2}, 1; x_1,x_2,x_3\bigg) \cr
    &\qquad\quad + \beta_k^2(3\beta_k^2-2\beta_i^2)\, 
     F_D^{(3)}\!\bigg(\frac{1}{2}, \frac{-1}{2}, \frac{-1}{2}, \frac{1}{2}, 1; 
       x_1,x_2,x_3\bigg) - 2\beta_k^2(\beta_i-\beta_k)^2\,
     F_D^{(3)}\!\bigg(\frac{1}{2}, \frac{-1}{2}, \frac{-1}{2}, \frac{3}{2}, 1; 
       x_1,x_2,x_3\bigg) \cr
    &\qquad\quad + \frac{(\beta_i-\beta_k)^4}{2}\,
         F_D^{(3)}\!\bigg(\frac{1}{2}, \frac{-1}{2}, \frac{-1}{2}, \frac{5}{2}, 
           1; x_1,x_2,x_3\bigg) \bigg]. \label{F4}
  \end{align}
\end{widetext}
\end{subequations}
We shall exploit these lengthy formulae in the following Section, and give 
alternative derivations of the limiting cases $\beta_i=\beta_k=1$ (resp.\ 
$\beta_i=1$, $\beta_k=0$) in Appendix~\ref{s:massless} (resp.\ \ref{s:Lorentz}).
Before that, let us make two remarks.

First, it is actually possible to derive shorter expressions, if instead of 
rewriting the polynomial ${\cal P}_n(u)$ as we did here, one rather looks for 
its zeros, factorizes the polynomial, and inserts the factorized ${\cal P}_n(u)$
under the square root, which yields $n$ extra terms in the numerator of the 
rational fraction. 
The resulting integral can then be expressed as a single Lauricella 
hypergeometric function of 5 (resp.\ 7) variables for $n=2$ (resp.\ $n=4$). 
This is indeed more compact, yet much more expensive to compute, which explains 
our choice. 

The second remark is that Eqs.~\eqref{F2}--\eqref{F4} may not always be the most
tractable expressions in practice, in particular when we shall be investigating the behavior
of the ${\cal F}_n$ in the limiting cases $\beta_i\ll 1$ or $\beta_i=1$.
As a matter of fact, they involve triply infinite series of variables that do 
not scale as a simple power of the velocities.
It is then simpler to work at the level of Eq.~\eqref{limvppp} multiplied with 
the overall prefactor from the first line of Eq.~\eqref{int_t(kernel)} and with 
the relative velocity --- performing e.g.\ the Taylor expansion in $\beta_i$ or 
considering the limit $\beta_i=1$ ---, and then to integrate over $\delta$.

\section{Results}
\label{s:results}

Let us now exploit the formulae~\eqref{limvp_0}--\eqref{def-F} derived in the 
previous Section.

\subsection{Eccentricity and velocity dependence}
\label{s:v-dependence}

In the previous Section, we have derived analytical formulae for the anisotropy
of the velocity distribution, which integrated over $|{\bf p}_k|$ yields the 
numerator in the definition~\eqref{vn} of $v_n(p_i)$, while the denominator is 
simply given by Eq.~\eqref{denom-vn}:
\begin{align}
\label{vn(pi)}
v_n(p_i) = {\cal N}_n{\cal K}_n(\epsilon)\!\int\!\dd p_k\,p_k\,N_k
  \tilde{f}_k(p_k)\,{\cal F}_n(\beta_i,\beta_k).
\end{align}
Before we perform this last integration, which requires some ansatz for the 
initial momentum distribution, we can analyze some properties of the functions 
${\cal K}_n$ and ${\cal F}_n$. 

First, if one fixes the differential cross section to enforce at most one 
collision in average per particle, i.e. according to the discussion below
\eqref{limvp_0} if $\sigma_{\rm d} = \kappa\sqrt{\pi}R/N_k$ with 
$\kappa\lesssim 1$, then the prefactors ${\cal N}_n$ simplify, and the whole 
dependence of $v_n(p_i)$ on $N_i$, $N_k$, $R$ cancels 
out. 

Next, the dependence on eccentricity is entirely given by the functions 
${\cal K}_n$, Eqs.~\eqref{K2}--\eqref{K4}.
That is, $v_2(p_i)$ scales like $\epsilon$ for small eccentricities, with less 
than 5\% deviation from the linear scaling up to $\epsilon=0.75$. 
$v_4(p_i)$ scales like $\epsilon^2$ for small eccentricities, with less than 5\%
deviation from the quadratic behavior up to $\epsilon=0.45$.
For larger eccentricities, ${\cal K}_2$ and ${\cal K}_4$ fall off to reach 0 
for $\epsilon=1$. 

An important fact is that the flow harmonics~\eqref{vn(pi)} depend on the 
particle velocity $\beta_i$, rather than on its momentum $p_i$. 
This clearly follows from the fact that the Boltzmann equation involves the 
velocities, not the momenta. 
As a consequence, particles with different masses but with the same velocity 
will have the same anisotropic flow $v_n(\beta_i)$: when considered as a 
function of momentum, this gives rise to mass ordering of the flow coefficients,
as is observed in experimental measurements~\cite{Adler:2003kt,Adams:2004bi}. 
For ``slow'' particles in the non-relativistic regime, $\beta_i\simeq p_i/mc$, 
so that in our model we find that the flow coefficients should coincide at the 
same $p_i/m_i$ for different particle species. 
This is noticeably what was found for slow particles emerging from an ideal 
expanding fluid~\cite{Borghini:2005kd}, that is for particles that underwent 
many rescatterings.

\begin{figure*}[t!]
\includegraphics*[width=0.49\linewidth]{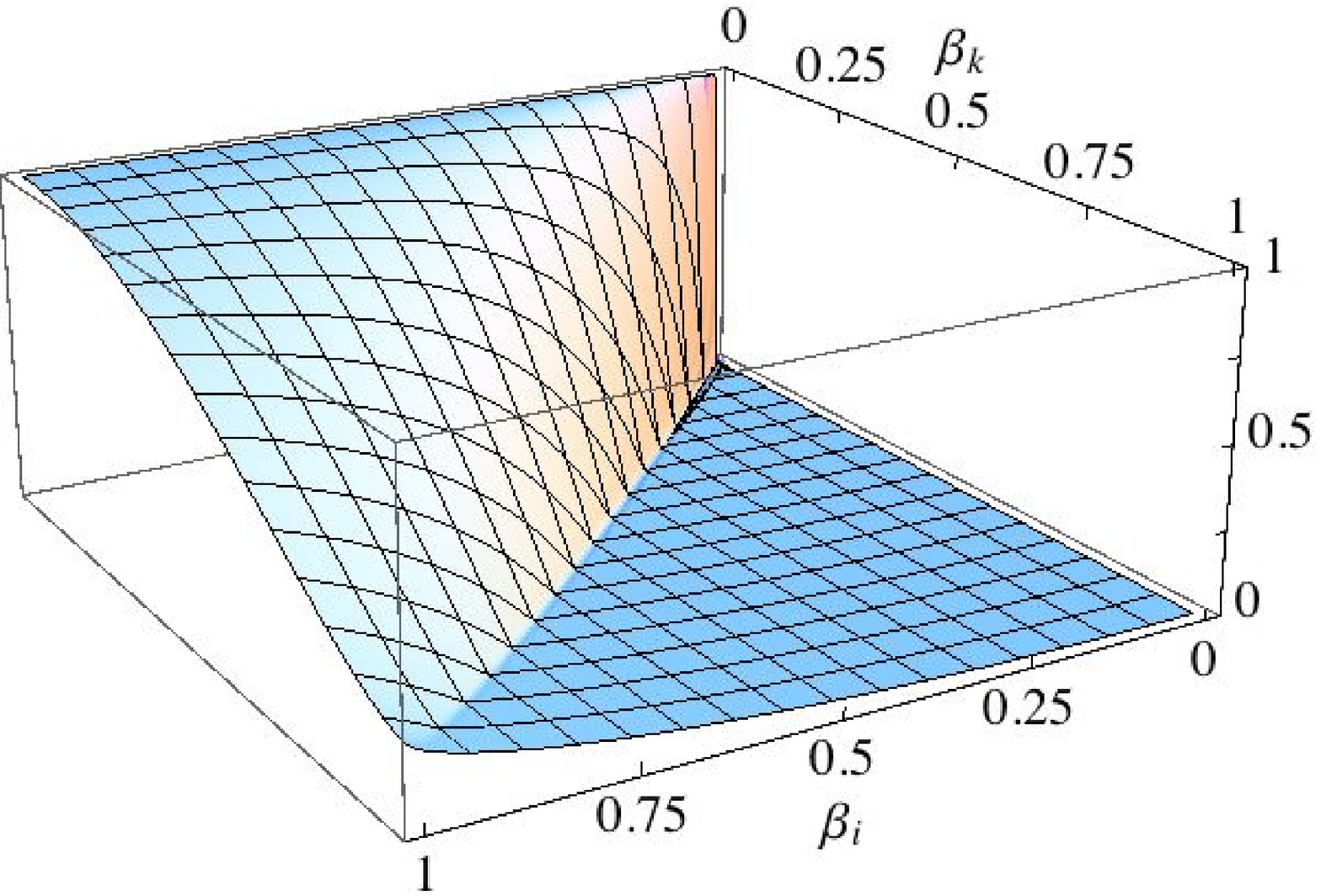}\hfill%
\includegraphics*[width=0.49\linewidth]{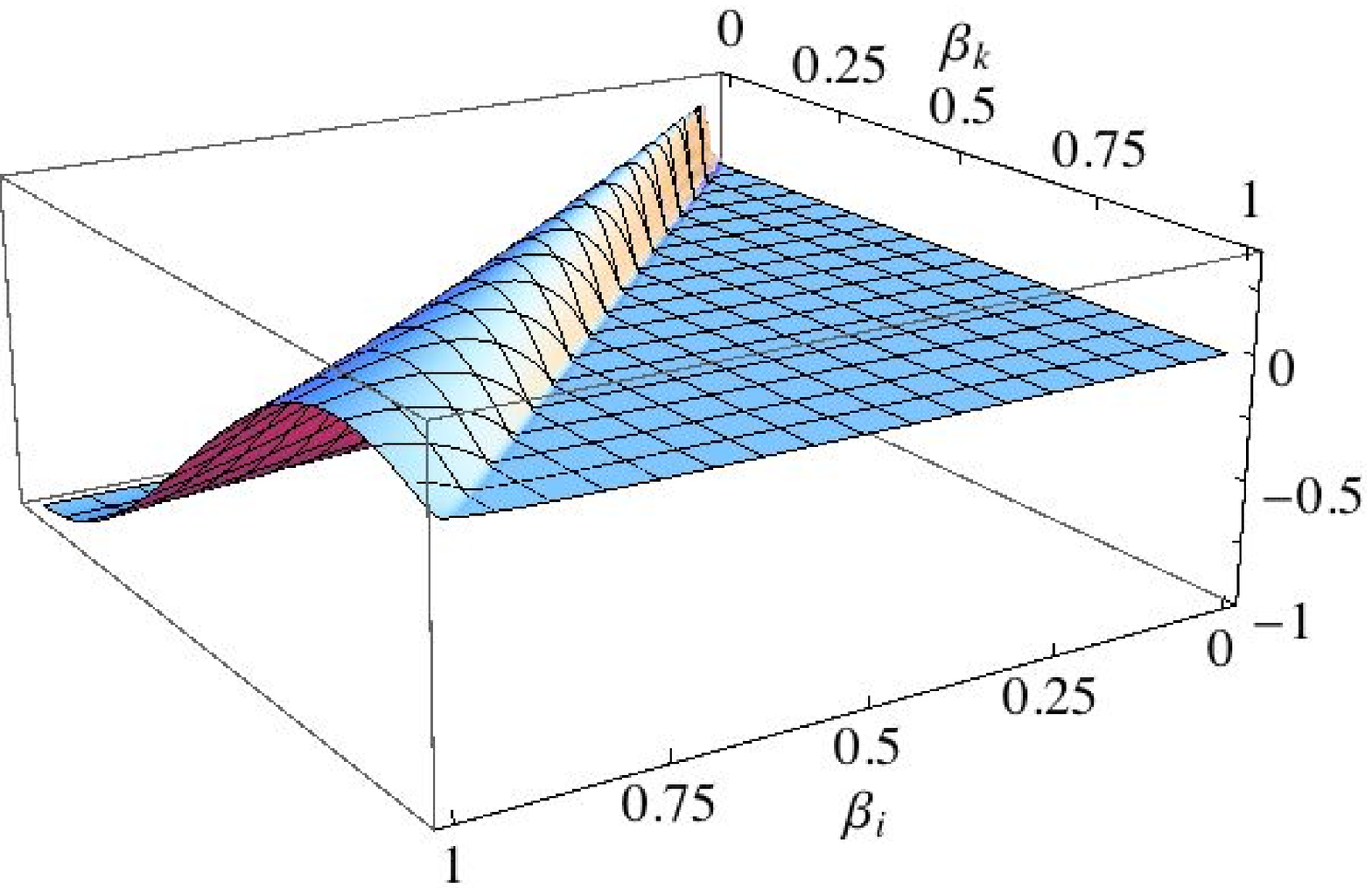}
\caption{\label{fig:plotsFn}Dependence of ${\cal F}_2$ (left) and $-{\cal F}_4$ 
  (right) on the velocities $\beta_i$, $\beta_k$.}
\end{figure*}

Let us discuss the dependence of ${\cal F}_n$ on the velocity $\beta_i$. 
At small $\beta_i$ and non-vanishing $\beta_k$,\footnote{For $\beta_k=0$, which
  implies $x_1=x_2=x_3=0$ and corresponds to the scattering of particles of 
  type $i$ on {\em fixed\/} centers of type $k$, the velocity-dependent parts 
  in Eqs.~\eqref{F2}--\eqref{F4} yield 1 for $\beta_i\neq 0$.
  The limit $\beta_i=\beta_k=0$ is thus singular, yielding different results 
  according to the order of the limits.
  This is rather obvious from the physical point of view, since it corresponds 
  to both particles staying at rest.} 
one finds the behaviors
\begin{equation}
\label{Fn(b_i<<1)}
{\cal F}_2(\beta_i,\beta_k) \sim \frac{\beta_i^2}{8}, \qquad
{\cal F}_4(\beta_i,\beta_k) \sim -\frac{\beta_i^4}{128}.
\end{equation}
These scalings will not be modified in the integration over $|{\bf p}_k|$. 
Since a small $\beta_i$ means a momentum $p_i\sim mc\beta_i$, one thus finds 
that in the region where particles $i$ are non-relativistic, 
\begin{equation}
v_n(p_i) \propto p_i^n,
\end{equation}
which is the expected behavior when the particle momentum distribution is smooth
at ${\bf p}={\bf 0}$~\cite{Danielewicz:1994nb}.

We show ${\cal F}_2$ and $-{\cal F}_4$ (to take into account the minus sign of 
${\cal N}_4$) as functions of $\beta_i$, $\beta_k$ in Fig.~\ref{fig:plotsFn}.
At fixed $\beta_k$, one finds as expected the scalings~\eqref{Fn(b_i<<1)} at 
small $\beta_i$ --- the factor 1/128 for ${\cal F}_4$ makes it difficult to 
display the quartic behavior, yet it is present.
These smooth behaviors persist up to $\beta_i=\beta_k$, i.e.\ as long as the 
``diffusing particles'' $i$ do not catch up with the ``scattering centers'' $k$ 
emitted in the same direction.

When $\beta_i\geq\beta_k$, the diffusing particles travel fast enough to probe 
the whole geometry of the distribution of scattering centers. 
There is thus for $\beta_i=\beta_k$ a transition from a regime where anisotropic
flow is built up ``locally'', involving only the diffusing particles that were 
close enough to the edge of the interaction region to find quickly a scattering 
center, to a ``global'' regime where every diffusing particle has a chance to 
scatter on an initially distant center.
As a consequence, the functions ${\cal F}_2$ and ${\cal F}_4$ vary sharply for 
$\beta_i\gtrsim\beta_k$.

${\cal F}_4$ is especially interesting, for it changes sign. 
When the scattering centers barely move, $\beta_k\ll\beta_i$, then the diffusing
particles quickly probe the initial Gaussian distribution~\eqref{dist(t=0)}. 
Now, a property of the latter is that 
\[
\frac{\langle r^2\cos 2\varphi_r \rangle}{\langle r^2\rangle} = -\epsilon,\qquad
\frac{\langle r^4\cos 4\varphi_r \rangle}{\langle r^4\rangle} = 
  \frac{3\epsilon^2}{2+\epsilon^2}, 
\]
where ($r, \varphi_r$) denote the polar coordinates of a point in position 
space: $x=r\cos\varphi_r$, $y=r\sin\varphi_r$.
One conventionally adds a global minus sign, to obtain a positive elliptic flow 
for $\epsilon>0$  [see Eq.~\eqref{eccentricity}]. 
In turn, this means that one would naturally expect that a Gaussian distribution
gives rise to a negative $v_4$.
This is exactly what comes out for $-{\cal F}_4$, which becomes negative when 
$\beta_i$ is significantly larger than $\beta_k$.

The flow harmonics at a given $\beta_i$ follow from integrating ${\cal F}_n$ 
over the whole $\beta_k$ range, Eq.~\eqref{vn(pi)}.
Depending on the shape of the spectrum $\tilde{f}_k(p_k)$, more weight is put
on different $\beta_k$ regions, resulting in different values of $v_n(p_i)$. 
This dependence on the spectrum will be particularly marked for $v_4(p_i)$ due 
to the non-monotonic shape of $-{\cal F}_4(\beta_i,\beta_k)$ at fixed $\beta_i$.

Let us finally discuss the behavior of the flow coefficients in limiting regimes
accessible in our approach. 
Corresponding direct computations, which are significantly simpler than going 
through Section~\ref{s:computations} and then taking the relevant limits, can be
found in Appendices~\ref{s:massless} and  \ref{s:Lorentz}.

The first limiting regime corresponds to the massless gas limit, in which both 
$\beta_i$ and $\beta_k$ are uniformly equal to $1$ (Appendix~\ref{s:massless}). 
The flow coefficients show an universal behavior, i.e.\ $v_n$ is independent of 
the assumed initial spectrum. 
When the cross section is fixed so that particles rescatter at most once in 
average (and exactly once in average for central collisions), we find
$v_2/\epsilon\simeq 0.083$ with less than 5\% deviation up to $\epsilon=0.75$.

The other limiting regime corresponds to that of a Lorentz gas of massless 
diffusing particles scattering on infinitely massive fixed centers, that is 
$\beta_i=1$, $\beta_k=0$ (Appendix~\ref{s:Lorentz}).
In that case the diffusing particles probe most cleanly the Gaussian 
distribution of scattering centers, and accordingly one finds a positive $v_2$ 
and a negative $v_4$, again universally independent of the initial momentum 
distribution, in perfect agreement with the initial geometry. 
If one requires a single rescattering per particle in average for $\epsilon=0$,
then $v_2\simeq 0.25\epsilon$, $v_4\simeq -0.094\epsilon^2$, that is 
$v_4/v_2^2\simeq -1.5$, across a wide $\epsilon$ range.

\subsection{Momentum dependence}
\label{s:p-dependence}

\begin{figure*}[ht!]
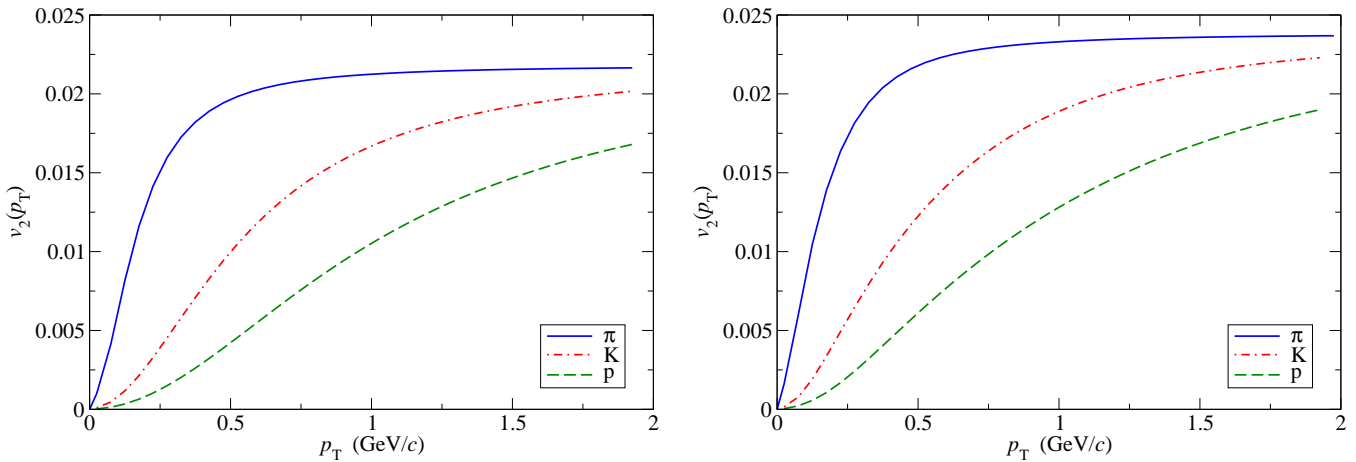

\includegraphics*[width=0.49\linewidth]{v2thermique}\hfill%
\includegraphics*[width=0.49\linewidth]{v2powerlaw}
\caption{\label{fig:v2}Elliptic flow $v_2(p_i)$ of pions (full line), kaons 
  (dot-dashed line) and protons (dashed line) assuming a thermal-like initial 
  spectrum~\eqref{thermal} (left) or a power-law spectrum~\eqref{power-law} 
  (right), for an eccentricity $\epsilon=0.1$.}
\end{figure*}
In order to present the flow coefficients as a function of momentum, as is 
customary, we now have to assume some initial spectrum $\tilde{f}_k(p_k)$ to 
insert in the integral~\eqref{vn(pi)}.
We choose two different distributions:
\begin{itemize}
\item A thermal-like spectrum
  \begin{equation}
  \label{thermal}
  \tilde{f}_k(p_k) = \frac{{\rm e}^{m_k c/T}}{T(m_k c+T)}\,
    {\rm e}^{-\sqrt{p_k^2+m_k^2c^2}/T},
  \end{equation}
  where we fix the inverse slope parameter $T$ so as to give pions an average 
  (transverse) momentum $\langle p_\pi\rangle=420$~MeV$/c$.

\item A QCD-inspired power-law spectrum, regulated at the origin
  \begin{equation}
  \label{power-law}
  \tilde{f}_k(p_k)=\frac{4}{\pi p_0^2}\frac{1}{1+\big(\frac{p_k}{p_0}\big)^4},
  \end{equation}
  where the characteristic momentum scale $p_0$ is fixed such that the average
  momentum per particle equals $420$~MeV$/c$.
\end{itemize}
Integrating Eq.~\eqref{vn(pi)} with these spectra necessitates a numerical 
approach. 

\begin{figure*}[t!]
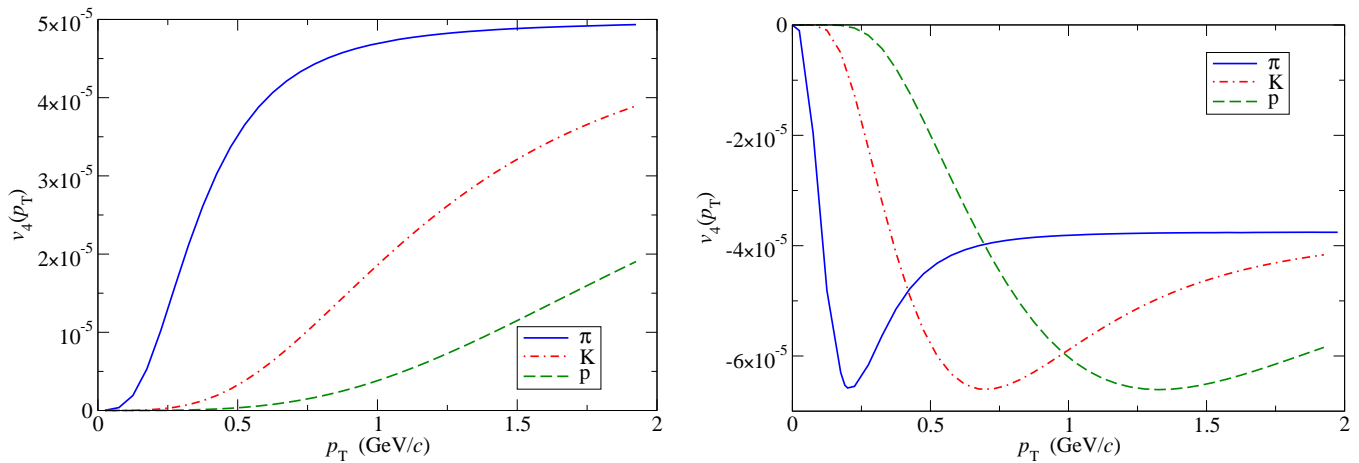

\includegraphics*[width=0.49\linewidth]{v4thermique}\hfill%
\includegraphics*[width=0.49\linewidth]{v4powerlaw}
\caption{\label{fig:v4}Hexadecupolar flow $v_4(p_i)$ of pions (full line), kaons
  (dot-dashed line) and protons (dashed line) assuming a thermal-like initial 
  spectrum~\eqref{thermal} (left) or a power-law spectrum~\eqref{power-law} 
  (right).}
\end{figure*}
To emphasize the mass ordering which we discussed above, we consider a mixture 
of three components with different masses (and different yields, to mimic a 
realistic situation, although we did not attempt to take experimental values): 
$80\%$ pions, $12.5\%$ kaons and $7.5\%$ protons.
Accordingly, the mixture is described by a system  of three coupled Boltzmann 
equations~\eqref{Boltzmann}, which all in all involve 6 different collision 
terms ($\pi$-$\pi$, $\pi$-K, $\pi$-p, K-K, K-p and p-p) that have to be taken 
into account carefully to ensure a fixed mean number --- which will be taken 
equal to 1 for $\epsilon=0$ --- of collisions per particle.
In turn, the flow harmonic $v_n(p_i)$ for a given species involves three 
different terms. 

To keep the number of free parameters to a minimum, we assume the same 
differential cross section $\sigma_{\rm d}$ for all these collisions, and the 
same inverse slope parameter $T$ or momentum scale $p_0$ for all species.
Note that this implies that the average momentum of heavier particles is not 
the same for both choices of spectra: for the power-law spectrum, all species 
have a mean momentum of $420$~MeV$/c$, while this only holds for pions for the 
thermal-like spectrum.

We display our results for $v_2(p_i)$ in Fig.~\ref{fig:v2} for both choices of 
initial spectrum, in the case $\epsilon=0.1$.
As anticipated, there is for both initial momentum distributions a clear 
mass-ordering of $v_2(p_i)$, which is thus no hallmark of the presence of a 
thermal system. 

Another striking feature is the size of the elliptic flow, of the order of a few
percent, i.e.\ $v_2/\epsilon$ of the order of 0.1--0.2 for both spectra.
Note that the magnitude of $v_2$ for a given species depends on its abundance, 
since we enforce the condition of an average single collision per particle on 
all particles, not for each species individually. 
Then, the more abundant a species, the more collisions it undergoes and thus 
the larger its elliptic flow.
For that reason, the absolute value of $v_2$ for a given initial momentum 
distribution is not very meaningful in our eyes --- it depends on the particle 
abundance, and additionally it is directly proportional to $\sigma_{\rm d}$, 
thus determined by our requiring a given average number of rescatterings per 
particle, which is an arbitrary choice of ours. 

Eventually, one can recognize the quadratic growth at small $p_i$ --- or rather,
at small $p_i/m_i$ --- for kaons and protons.
For pions, the non-relativistic regime where this scaling holds is not visible. 

\begin{figure}[t!]
\includegraphics*[width=0.98\linewidth]{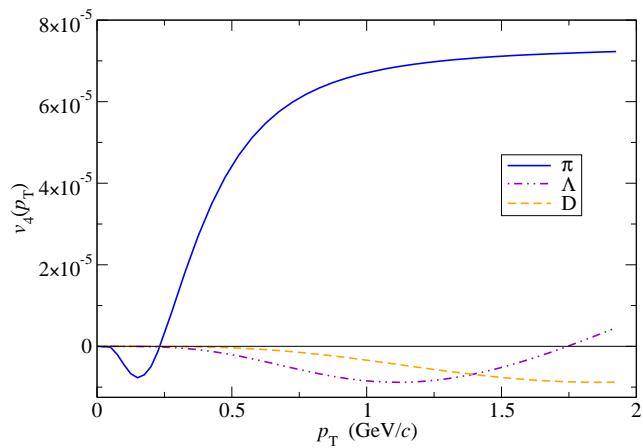}
\caption{\label{fig:v4b}Hexadecupolar flow $v_4(p_i)$ of  pions (full line), 
  $\Lambda$ (dot-dashed line) and D-mesons (dashed line), assuming a 
  thermal-like initial spectrum~\eqref{thermal}.}
\end{figure}
As explained in the discussion of ${\cal F}_4$, we expect a richer variety of 
behaviors for the fourth harmonic $v_4(p_i)$. 
Accordingly, we present more plots.
In Fig.~\ref{fig:v4} we show $v_4(p_i)$ for pions, kaons and protons for both 
initial momentum distributions: thermal-like on the left panel and power-law on
the right panel. 
Figure~\ref{fig:v4b} then shows $v_4(p_i)$ for a mixture of 88\% pions, 8\% 
$\Lambda$ and 4\% D-mesons, with a thermal-like initial spectrum, so as to 
have even heavier particles, with smaller $\beta_k$ at a fixed $p_k$.

Again, $v_4(p_i)$ exhibits mass ordering in the low momentum, non-relativistic 
region, irrespective of the choice of spectrum.
A further consequence of the dependence of $v_4$ on $\beta_i$ rather than on 
$p_i$ is mostly visible on the right panel of Fig.~\ref{fig:v4}, although 
present on all our plots of elliptic and hexadecupolar flow. 
Here, one clearly sees that the proton and kaon $v_4$ have exactly the same 
shape as the pion one, although dilated along the momentum axis; in particular,
they all reach the same minimum value. 

The most striking feature when comparing both panels of Fig.~\ref{fig:v4} is 
obviously the change of sign of $v_4(p_i)$ according to the choice of spectrum.
As explained above, this comes from weighting the various velocity regions of 
${\cal F}_4(\beta_i,\beta_k)$ differently in the integral over $|{\bf p}_k|$:
the power-law spectrum weights the low-velocity region more than the 
thermal-like distribution, so that it emphasizes regions of negative 
$-{\cal F}_4$ more, which explains the negative $v_4(p_i)$. 
Note that these regions of negative $-{\cal F}_4$ corresponds to where 
${\cal F}_2$ is largest, which agrees with the larger $v_2(p_i)$ found with the 
power-law spectrum.

When one replaces kaons and protons by smaller amounts of the significantly 
heavier $\Lambda$ and D, there are more ``slow'' scattering centers in the 
system, i.e.\ one again puts more emphasis on the negative $-{\cal F}_4$ 
regions.
As a result, the pions $v_4(p_\pi)$ for a thermal-like spectrum now takes 
negative in a small $p_\pi$ region, before turning positive again 
(Fig.~\ref{fig:v4b}).
A small region in $p_\pi$ amounts to much larger regions in $p_\Lambda$ or 
$p_{\rm D}$, so that $v_4$ for those species remains negative over most of the 
displayed range.\footnote{Interestingly, the STAR Collaboration has reported 
  values of $v_4$ for antiprotons and $\Lambda$ which are compatible with 0 or 
  even negative in the low momentum region~\cite[Fig.~7]{Abelev:2007qg}.}

For the sake of completeness, let us mention the ratio $v_4(p_i)/v_2(p_i)^2$. 
We do not show our results for this ratio because we do not give them too much 
worth: in our model $v_4(p_i)/v_2(p_i)^2$ obviously scales as $1/\sigma_{\rm d}$
and might not be reliable in the limit of a low number of collisions we are
considering, as we have not estimated the size of the corrections to our 
results.

\section{Discussion}
\label{s:discussion}

Before we start discussing our results, let us summarize the salient features 
of our study. 
We have solved the Boltzmann equation in the limit of a (very) small average 
number of rescatterings per outgoing particle, which allows us to compute the 
anisotropic flow harmonics in that limit. 
We then find that sizable elliptic flow is generated already after a single 
rescattering and that $v_2(p_i)$ exhibits mass ordering. 
Additionally, we find that the fourth harmonic $v_4(p_i)$ is quite sensitive to
the initial momentum distribution of particles. 

Our approach can only be considered as a toy model for the phenomenology of 
nucleus--nucleus collisions, so that one should only retain the qualitative 
aspects, and not attach too much worth to the quantitative values we obtain. 
A first reason is that we have deliberately reduced the numbers of ingredients 
and parameters to a minimum: Gaussian initial shape, hard-spheres-type universal
isotropic cross section, and initial spectra depending on one parameter only.
The second reason is that we have restricted ourselves to the leading term in an
expansion in powers of the cross section of the solution of the Boltzmann 
equation.
If one wants to evaluate the quantitative uncertainties on our results, one has
to explicitly compute the next term in the expansion, or at least to estimate 
it.
This can be done, at least numerically, yet we have not attempted to do it. 

Nonetheless, we feel confident that the qualitative trends we have found are 
quite robust, and must also be present in more precise studies. 
In particular, the mass ordering of $v_n(p_T)$ follows from the fact that the 
Boltzmann equation framework only involves the velocities of particles, and thus
its origin appears to be intrinsic to the kinetic equation.

Let us eventually discuss possible generalizations of the present work. 
First, one may try different shapes for the initial distribution in position 
space, in particular more realistic, finite shapes.

A desirable improvement that might become feasible would be the possibility to 
perform both angular integrals first, so as to isolate the time evolution of the
flow harmonics --- as is the case for the Lorentz gas, see Appendix~\ref{%
  s:Lorentz}.
This would help implement our study as a model for two rather different cases: 
first, as a description of the very first steps of the expansion, which would 
provide initial conditions for a subsequent hydrodynamic calculation.
Second, one may possibly describe momentum anisotropy at high momentum: 
since we only consider the loss term of the (elastic) collision term, our 
approach is not far from constituting a phenomenological description of the loss
of partons with a given high momentum due to their branching in two lower 
momentum partons.

Another possible generalization consists in introducing some dependences in the 
cross section: either on the collision type, or on the center-of-mass energy of
the rescatterings, or allowing for anisotropic cross sections. 

Finally, taking care of the next order(s) in the expansion will be needed, if 
only to estimate the uncertainty on the results. 

\begin{acknowledgments}
  We thank Jean-Paul Blaizot and Jean-Yves Ollitrault for discussions.
  C.~G.\ acknowledges support from the Deutsche Forschungsgemeinschaft under  
  grant GRK 881.
\end{acknowledgments}

\appendix

\section{Lauricella's multivariate hypergeometric function $F_D$}
\label{s:Lauricella}

In this Appendix we present some basic properties of the function $F_D^{(3)}$ 
with the help of which we express our analytical results for the anisotropy of 
the velocity distribution.
 
The Lauricella function of the fourth kind of $m\in\mathbb{N}^+$ variables 
$z_1,\ldots,z_m$ is defined by~\cite{Lauricella:1893}
\begin{align}
\label{def_FD}
F_D^{(m)}(a,b_1,\ldots,b_m,c;z_1,\ldots,z_m) =\qquad\quad &\cr
  \sum_{k_1,\ldots,k_m=0}^\infty\!\!\!
  \frac{(a)_{k_1+\cdots+k_m} (b_1)_{k_1}\cdots(b_m)_{k_m}}{%
    (c)_{k_1+\cdots+k_m} k_1!\cdots k_m!}z_1^{k_1}&\cdots z_m^{k_m},
  \qquad\ 
\end{align}
where the Pochhammer symbol $(a)_k$ is defined by
\[
(a)_k = \frac{\Gamma(a+k)}{\Gamma(a)}.
\]
The series converges absolutely for $|z_1|<1$, \ldots, $|z_m|<1$.
For $m=1$, it reduces the Gauss hypergeometric function ${}_2F_1(a,b,c,z)$, and 
for $m=2$, to the Appell function $F_1$.

If the real parts of $a$ and $c-a$ are positive, the function admits the 
integral representation~\cite{Lauricella:1893}
\begin{align}
\label{FD_as_int}
F_D^{(m)}(a,b_1,\ldots,b_m,&c; z_1,\ldots,z_m) = \cr
  \frac{\Gamma(c)}{\Gamma(a)\,\Gamma(c-a)}& \!\int_0^1\!
  \frac{u^{a-1}u^{c-a-1}\,\dd u}{(1-z_1 u)^{b_1}\cdots(1-z_m u)^{b_m}}. \qquad
\end{align}
This representation allows the analytical continuation of $F_D^{(m)}$ to the cut
complex $z_i$-planes deprived of the half-line from 1 to infinity along the 
positive real axis.

If ${\rm Re}\,(c-a-b_1-\cdots-b_m)>0$ one finds
\begin{align}
\label{Lauricella(1)}
F_D^{(m)}(a,b_1,\ldots,&\,b_m,c; 1,\ldots,1) = \cr
  &\frac{\Gamma(c)\,\Gamma(c-a-b_1-\cdots-b_m)}{%
    \Gamma(c-a)\,\Gamma(c-b_1-\cdots-b_m)}.\quad
\end{align}

\section{Massless diffusing particles and scattering centers}
\label{s:massless}

In this Appendix we calculate the anisotropies of the velocity distribution and 
the average number of collisions in the case where both colliding particle 
species are massless, which leads to $\beta_i=\beta_k=1$.  
We assume that the particles remain distinguishable. 

In that case, the relative velocity~\eqref{v_ik(mass)} reduces to
\begin{equation}
\label{v_ik(m=0)}
v_{ik} = c \big[ 1 - \cos(\varphi_i-\varphi_k) \big] = 
  2c\sin^2\frac{\varphi_i-\varphi_k}{2},
\end{equation}
and the kernel~\eqref{kernel} becomes, after performing the change of angular 
variables~\eqref{phi-vs-alpha,delta},
\begin{align}
\label{kernel(m=0)}
{\cal C}(t,{\bf p}_i,{\bf p}_k) =& 
  -\frac{N_iN_k\sigma_{\rm d}c\sqrt{1-\epsilon^2}}{4\pi^2R^2}\,\tilde{f}_i(p_i)
    \,\tilde{f}_k(p_k)\, \cr
 & \times \sin^2\delta\,\exp\!\bigg[\!-\frac{c^2t^2}{R^2}
   \big( 1-\epsilon\cos 2\phi^{\rm p} \big)\sin^2\delta \bigg],\cr
\end{align}
which simplifies significantly the ensuing calculations and the final formulae
compared to the case of arbitrary velocities.

As above, we need to integrate the kernel~\eqref{kernel(m=0)} multiplied by 
$\cos n(\phi^{\rm p}+\delta)$ over both angles $\phi^{\rm p}$ and $\delta$ and 
over time to obtain the average number of collisions ($n=0$) and the flow 
coefficients. 

The integral over time is trivial and yields
\[
\label{int_t(kernel,m=0)}
-\frac{N_iN_k\sigma_{\rm d}\sqrt{1-\epsilon^2}}{8\pi^{3/2}R}\,\tilde{f}_i(p_i)\,
  \tilde{f}_k(p_k)\,\frac{|\sin\delta|\,\cos n(\phi^{\rm p}+\delta)}{%
    \sqrt{1-\epsilon\cos 2\phi^{\rm p}}}.
\]
The angular part of this expression should be compared with 
Eq.~\eqref{limvppp}.

Integrating over $\delta\in[-\pi,\pi]$ is also straightforward and gives
\begin{equation}
\label{drosophile}
\frac{N_iN_k\sigma_{\rm d}\sqrt{1-\epsilon^2}}{2(n^2-1)\pi^{3/2}R}\,
  \tilde{f}_i(p_i)\,\tilde{f}_k(p_k)\,
  \frac{\cos n\phi^{\rm p}}{\sqrt{1-\epsilon\cos 2\phi^{\rm p}}}.
\end{equation}

There remains the integral over $\phi^{\rm p}\in[-\pi,\pi]$.
One finds for $n=0$ 
\begin{equation}
\label{limvp_0(m=0)}
\frac{2N_iN_k\sigma_{\rm d}}{\pi^{3/2}R}\tilde{f}_i(p_i)\,\tilde{f}_k(p_k)\,
  \sqrt{1-\epsilon}\,K\!\bigg(\!\sqrt{\dfrac{2\epsilon}{1+\epsilon}}\bigg),
\end{equation}
where we have dropped the minus sign, which comes from our integrating the loss 
term of the Boltzmann equation, and is irrelevant for the computation of the 
mean number of collisions. 
Now, if one takes $\beta_i=\beta_k=1$ in Eq.~\eqref{limvp_0}, then 
$x_1=x_2=x_3=1$ and the value of Lauricella function in the second line is given
by Eq.~\eqref{Lauricella(1)} with $a=\frac{1}{2}$, $b_1=b_2=-\frac{1}{2}$, 
$b_3=\frac{1}{2}$, $c=1$, i.e.\ it equals $2/\pi$, so that Eq.~\eqref{%
  limvp_0(m=0)} does represent the limit of Eq.~\eqref{limvp_0}.

Integrating Eq.~\eqref{limvp_0(m=0)} over $|{\bf p}_i|$ and $|{\bf p}_k|$ is 
straightforward and one finds that the total number of collisions between 
particles $i$ and $k$ is
\[
N_{\rm coll} = \frac{2N_iN_k\sigma_{\rm d}}{\pi^{3/2}R}\,\sqrt{1-\epsilon}\,
  K\!\bigg(\!\sqrt{\dfrac{2\epsilon}{1+\epsilon}}\bigg).
\]
If one requires that each ``diffusing particle'' $i$ undergo in average one 
collision, so that $N_{\rm coll} = N_i$, for the most central collisions, where 
the eccentricity-dependent term equals $\pi/2$, then this fixes the 
differential cross section to 
\begin{equation}
\label{sigma(1-coll)}
\sigma_{\rm d}^{\rm 1\,coll} = \frac{\sqrt{\pi}}{N_k}R.
\end{equation}

Let us now turn to the anisotropic flow harmonics. 
The integral of Eq.~\eqref{drosophile} over $\phi^{\rm p}$ for $n=2$ or $n=4$ 
reads after some algebra
\begin{subequations}
\label{mouche}
  \begin{align}
  \frac{N_iN_k\sigma_{\rm d}}{12\sqrt{\pi}R}\tilde{f}_i(p_i)\,\tilde{f}_k(p_k)&
    \sqrt{1\!-\!\epsilon^2}\,
    {}_2F_1\!\bigg(\frac{3}{4},\frac{5}{4};2;\epsilon^2\bigg) \epsilon, \\
  \frac{N_iN_k\sigma_{\rm d}}{160\sqrt{\pi}R}\tilde{f}_i(p_i)\,\tilde{f}_k(p_k)&
    \sqrt{1\!-\!\epsilon^2}\,
    {}_2F_1\!\bigg(\frac{5}{4},\frac{7}{4};3;\epsilon^2\bigg) \epsilon^2.
  \end{align}
\end{subequations}
One recognizes the eccentricity-dependent~\eqref{K2}--\eqref{K4}.
Furthermore using Eq.~\eqref{Lauricella(1)} with $a=\frac{1}{2}$, 
$b_1=b_2=-\frac{1}{2}$, $c=1$ and $b_3=\frac{1}{2}$, $-\frac{1}{2}$ or 
$-\frac{3}{2}$, one finds ${\cal F}_2(1,1) = 2/3\pi$ and 
${\cal F}_4(1,1) = 2/15\pi$, so that Eqs.~\eqref{def-F} reduce to Eqs.~\eqref{%
  mouche} for $\beta_i=\beta_k=1$.

Again, the latter can be integrated over $|{\bf p}_k|$, divided by the 
denominator~\eqref{denom-vn} of the flow coefficient definition, which yields 
the flow harmonics
\begin{align*}
v_2(p_i) =& \frac{N_k\sigma_{\rm d}}{12\sqrt{\pi}R}\sqrt{1\!-\!\epsilon^2}\,
  {}_2F_1\!\bigg(\frac{3}{4},\frac{5}{4};2;\epsilon^2\bigg) \epsilon, \\
v_4(p_i) =& \frac{N_k\sigma_{\rm d}}{160\sqrt{\pi}R}\sqrt{1\!-\!\epsilon^2}\,
  {}_2F_1\!\bigg(\frac{5}{4},\frac{7}{4};3;\epsilon^2\bigg) \epsilon^2.
\end{align*}
Note that these values are actually independent of the momentum, and in 
particular do not vanish when $p_i$ goes to 0!
This comes from the facts that in our model the harmonics $v_n$ depend 
primarily on velocity, rather than on momentum, and that massless particles 
always travel with the velocity of light, irrespective of their momentum. 

If the cross section is fixed to its value~\eqref{sigma(1-coll)} to ensure one 
collision per particle in the most central collisions, these flow harmonics 
becomes
\begin{align*}
v_2(p_i) =& \ \frac{1}{12}\sqrt{1\!-\!\epsilon^2}\,
  {}_2F_1\!\bigg(\frac{3}{4},\frac{5}{4};2;\epsilon^2\bigg) \epsilon, \\
v_4(p_i) =& \ \frac{1}{160}\sqrt{1\!-\!\epsilon^2}\,
  {}_2F_1\!\bigg(\frac{5}{4},\frac{7}{4};3;\epsilon^2\bigg) \epsilon^2.
\end{align*}
This gives a rather large $v_2/\epsilon\simeq 0.08$ for $\epsilon$ up to 0.75, 
and a ratio $v_4/v_2^2$ that equals 0.9 for $\epsilon=0$ and slowly increases 
quadratically with $\epsilon$.

\section{Massless diffusing particles scattering on fixed centers}
\label{s:Lorentz}

We now consider the case of a Lorentz gas of massless diffusing particles 
($\beta_i=1$) scattering on fixed, infinitely massive centers ($\beta_k=0$).

In that case, the relative velocity~\eqref{v_ik(mass)} is $v_{ik}=c$, 
independent of the relative angle $\varphi_i-\varphi_k$, which is quite natural 
as $\varphi_k$ is undefined.
In turn, the kernel~\eqref{kernel} simplifies to
\begin{align*}
{\cal C}(t,{\bf p}_i,{\bf p}_k) =& 
  -\frac{N_iN_k\sigma_{\rm d}c\sqrt{1-\epsilon^2}}{8\pi^2R^2}\,\tilde{f}_i(p_i)
    \,\tilde{f}_k(p_k) \\
 & \times \exp\!\bigg[\!-\frac{c^2t^2}{4R^2}
   \big( 1+\epsilon\cos 2\varphi_i \big)\bigg], 
\end{align*}
independent of $\varphi_k$, over which one can at once integrate, which amounts
to multiplying the kernel with $2\pi$. 
The resulting expression, multiplied with $\cos n\varphi_i$, is easily 
integrated over $\varphi_i$, yielding
\begin{align}
\label{dvn/dt...almost}
(-1)^{n/2+1}&\frac{N_iN_k\sigma_{\rm d}c\sqrt{1-\epsilon^2}}{2R^2}\,
  \tilde{f}_i(p_i)\,\tilde{f}_k(p_k) \cr
 & \times {\rm e}^{-c^2t^2/4R^2}\,
  I_{\!\frac{n}{2}}\!\bigg( \frac{c^2t^2}{4R^2}\epsilon \bigg),
\end{align}
where we have assumed that $n$ is even, and where $I_{n/2}$ denotes the modified
Bessel function of the first kind of order $n/2$. 
Note that the factor $(-1)^{n/2+1}$ comes in part from the global minus sign due
to our considering the loss term of the Boltzmann equation, and in part from 
our writing the Bessel function with a positive argument.

Now, integrating Eq.~\eqref{dvn/dt...almost} over time and over $|{\bf p}_k|$ 
and dividing by the denominator~\eqref{denom-vn} yields the flow harmonic 
$v_n(p_i)$, that is the large-time limit of $v_n(p_i,t)$.
The integral over time of the second line of Eq.~\eqref{dvn/dt...almost} 
gives~\cite[formula 2.15.3.2]{Prudnikov2}
\begin{equation}
\label{belle_int}
\frac{R\sqrt{\pi}}{c}\frac{(n-1)!!}{2^{n+1}(\frac{n}{2})!}\epsilon^{n/2}\,
  {}_2F_1\!\bigg( \frac{n+1}{4},\frac{n+3}{4};\frac{n}{2}+1;\epsilon^2\bigg),
\end{equation}
with $(2n-1)!!=1\cdot3\ldots(2n-1)$ if $n\geq 1$, 1 if $n=0$.

Taking into account the prefactor from Eq.~\eqref{dvn/dt...almost} and 
performing the trivial integrals over $|{\bf p}_k|$ and $|{\bf p}_i|$ gives 
with $n=0$ the total number of collisions
\begin{align*}
N_{\rm coll} =&\ \frac{N_iN_k\sigma_{\rm d}\sqrt{\pi}}{2R} \sqrt{1-\epsilon^2}\,
  {}_2F_1\!\bigg( \frac{1}{4},\frac{3}{4};1;\epsilon^2\bigg) \\
 =& \ \frac{N_iN_k\sigma_{\rm d}}{\sqrt{\pi}R}\,\sqrt{1-\epsilon}\,
  K\!\bigg(\!\sqrt{\frac{2\epsilon}{1+\epsilon}}\bigg),
\end{align*}
which is indeed the limit of Eq.~\eqref{limvp_0} for $\beta_k=0$ --- i.e.\ 
$x_1=x_2=x_3=0$ --- after integrating out the initial spectra. 
To ensure one collision per diffusing particle for $\epsilon=0$, and thereby 
less than one collision over the whole eccentricity range, the cross section 
should be
\begin{equation}
\label{sigma(1-coll,Lorentz)}
\sigma_{\rm d}^{\rm 1\,coll} = \frac{2}{N_k\sqrt{\pi}}R.
\end{equation}

Using Eq.~\eqref{belle_int} with $n=2$ or $n=4$ and restoring the necessary 
prefactor from Eq.~\eqref{dvn/dt...almost}, one derives the flow harmonics
\begin{align}
v_2(p_i) = & \ \frac{N_k\sigma_{\rm d}\sqrt{\pi}}{8R}\,\sqrt{1-\epsilon^2}\,
  {}_2F_1\!\bigg( \frac{3}{4},\frac{5}{4};2;\epsilon^2\bigg) \epsilon, \\
v_4(p_i) = & -\frac{3N_k\sigma_{\rm d}\sqrt{\pi}}{64R}\,\sqrt{1-\epsilon^2}\,
  {}_2F_1\!\bigg( \frac{5}{4},\frac{7}{4};3;\epsilon^2\bigg)\epsilon^2, 
\end{align}
which is what Eqs.~\eqref{def-F} give for $\beta_k=0$, $\beta_i\neq 0$.
Replacing $\sigma_{\rm d}$ by its value~\eqref{sigma(1-coll,Lorentz)} gives
\begin{align*}
v_2(p_i) = & \ \frac{1}{4}\,\sqrt{1-\epsilon^2}\,
  {}_2F_1\!\bigg( \frac{3}{4},\frac{5}{4};2;\epsilon^2\bigg) \epsilon, \\
v_4(p_i) = & -\frac{3}{32}\,\sqrt{1-\epsilon^2}\,
  {}_2F_1\!\bigg( \frac{5}{4},\frac{7}{4};3;\epsilon^2\bigg)\epsilon^2.
\end{align*}
That is, the diffusion of massless particles on fixed scattering centers leads 
to a large, momentum independent $v_2/\epsilon=0.25$ for $\epsilon$ up to 
$0.75$, while $v_4$ is negative, which directly reflects the geometry of the 
distribution, as discussed in Section~\ref{s:p-dependence}.
This leads to a ratio $v_4/v_2^2=-1.5$.

If we do not integrate Eq.~\eqref{dvn/dt...almost} over time, but perform the 
integration over $|{\bf p}_k|$ and the division by $N_i\tilde{f}(p_i)$, we 
obtain the time derivative $\dd v_n/\dd t$:
\[
\frac{\dd v_n}{\dd t} = (-1)^{n/2+1}
  \frac{N_k\sigma_{\rm d}c\sqrt{1-\epsilon^2}}{R^2}\,{\rm e}^{-c^2t^2/4R^2}
  I_{\!\frac{n}{2}}\!\bigg( \frac{c^2t^2}{4R^2}\epsilon \bigg).
\]
This derivative is interesting, for it provides us at once with the small-time 
evolution of the flow coefficients, valid for $t\ll 2R/c$: 
\[
\frac{\dd v_n}{\dd t} \sim (-1)^{n/2+1}
  \frac{N_k\sigma_{\rm d}c\sqrt{1-\epsilon^2}}{(\frac{n}{2})!\,R^2}\,
  \bigg(\frac{ct\sqrt{\epsilon}}{4R}\bigg)^{\!\!n},
\]
i.e.\ after integration
\[
v_n(t) \propto (-1)^{n/2+1} t^{n+1}.
\]
One thus recovers the behavior found in Ref.~\cite{Heiselberg:1998es} and 
confirmed in Monte Carlo simulations~\cite{Gombeaud:2007ub}.
We were unfortunately not able to derive this result analytically in the more 
general of collisions of massive particles, by performing both angular 
integrals before that over time without making some additional assumption (as
e.g.\ an expansion in small eccentricities).

\end{document}